\documentclass[twocolumn,showpacs,preprintnumbers,amsmath,amssymb,superscriptaddress]{revtex4}

\usepackage{longtable}
\usepackage{graphicx}
\usepackage{dcolumn}
\usepackage{bm}

\begin{document}

\title{Prediction of Underground Argon Content for Dark Matter Experiments}
\newcommand{\usd}{Department of Physics, The University of South Dakota, Vermillion, South Dakota 57069}
\newcommand{\usdc}{Department of Chemistry, The University of South Dakota, Vermillion, South Dakota 57069}
\newcommand{\lanl}{Los Alamos National Laboratory, Los Alamos, NM 87545}
\newcommand{\hua}{Institute of Particle Physics, Huazhong Normal University, Wuhan 430079, China}
\author{     D.-M.~Mei          }\altaffiliation[Electronic Address: ]{dongming.mei@usd.edu}	\affiliation{	\usd 	}
\author{    Z.-B.~Yin         }\altaffiliation[Permanent Address: ]{\hua} \affiliation{	\usd	} \affiliation{	\hua	}
\author{J. Spaans		}\affiliation{	\usd	}  
\author{M. Koppang		}\affiliation{	\usdc	}  
\author{	A.~Hime		}\affiliation{	\lanl	}				
\author{	C. Keller	}\affiliation{	\usd	}
\author{V.M. Gehman               }\affiliation{  \lanl  }
\date{\today}

\begin{abstract}
In this paper, we demonstrate the use of physical models to evaluate the production of $^{39}$Ar and $^{40}$Ar underground. 
Considering both cosmogenic $^{39}$Ar production and radiogenic $^{40}$Ar production in situ and from external
sources, we can derive the ratio of $^{39}$Ar to $^{40}$Ar in underground sources. We show for the first time that 
the $^{39}$Ar production 
underground 
is dominated by stopping negative muon capture on $^{39}$K and ($\alpha,n)$ induced subsequent $^{39}$K(n,p)$^{39}$Ar 
reactions. 
The production of $^{39}$Ar is shown as a function of depth. We demonstrate that argon depleted in $^{39}$Ar can be
 obtained only if the depth of the underground resources is greater than 500 m.w.e. below the surface. Stopping negative 
muon capture on $^{39}$K dominates over radiogenic production at depths of less than 2000 m.w.e., and that production 
by muon-induced neutrons is subdominant at any depth. The depletion 
factor depends strongly on both radioactivity level and  potassium content in the rock. We measure the radioactivity
concentration and  potassium concentration in the rock for
 a potential site of an underground argon source in South Dakota. 
 Depending on the probability of $^{39}$Ar and $^{40}$Ar 
produced underground being dissolved in the water, the upper limit of the concentration of $^{39}$Ar in
 the underground water at this site 
is estimated to be in a range of a factor of 1.6 to 155 less 
than the $^{39}$Ar concentration in the atmosphere. 
 The calculation tools presented in this paper are also critical to the
dating method with $^{39}$Ar. 
  
\end{abstract}

\pacs{95.35.+d, 11.10.Lm, 29.40.Mc}
\maketitle

\section{Introduction}
Argon is expected to be an excellent target for direct dark matter detection experiments searching for Weakly Interacting 
Massive Particles 
(WIMPs)~\cite{bh, warp}. However, natural argon collected from the atmosphere contains $^{39}$Ar and $^{42}$Ar that are 
radioactive isotopes. 
The ratio of $^{39}$Ar to $^{40}$Ar is measured to be 8.1$\times$10$^{-16}$g/g~\cite{hhl,wku, wrap1} for atmospheric 
argon. Such a  concentration 
results in 1 Bq/kg of specific activity of $^{39}$Ar ($Q = 565$ keV, t$_{1/2}$= 269 y~\cite{sto}). 
Existing experimental limits on 
the $^{42}$Ar concentration 
in the atmosphere are less than 10$^{-18}$ $^{42}$Ar atoms per $^{40}$Ar atom~\cite{car,pce, asb}, which makes 
about 0.018 Hz/kg of specific 
activity of $^{42}$Ar ($Q= 3520$ keV and t$_{1/2}$ = 33 y). The production of $^{39}$Ar in the atmosphere is dominated by 
the $^{40}$Ar(n, 2n)
process~\cite{hhl} with a reaction threshold of 1011.86 keV. $^{42}$Ar is produced in the atmosphere
through $^{40}$Ar +n $\rightarrow$ $^{41}$Ar and $^{41}$Ar+n $\rightarrow$$^{42}$Ar. The neutrons in the above reactions 
are from 
cosmic rays.

Recently developed argon-based dark matter detectors measure scintillation light 
induced by low-energy nuclear recoils due to elastic scattering of WIMPs. 
 Argon has two distinct mechanisms for the emission of scintillation light: prompt light (6 ns) and delayed 
light (1.5 $\mu$s)~\cite{ahi}. 
The ratio of the prompt light to total light has been demonstrated experimentally to be well separated between
 electronic recoils and nuclear recoils~\cite{pbe, hugh}.  
This property allows these two types of events to be distinguished on an event by event basis. The discrimination power 
is proved to be up to $~10^{7}$ electronic recoils 
versus one nuclear recoil~\cite{pbe,hugh} at 20 keV electron-equivalent (ee) energy. Such a high 
pulse shape discrimination (PSD) power makes
argon a very attractive dark matter target for the planned dark matter program at DUSEL. However, the existence of $^{39}$Ar 
and $^{42}$Ar potentially limits the 
detection sensitivity of next generation DUSEL dark matter experiments with multi-ton target masses. For instance, argon 
extracted from air results in $^{39}$Ar beta-decay 
at a level of 1 Hz/kg~\cite{wrap1}; a 5-ton detector would give a trigger rate of 5000 Hz and a total of 
3.2 $\times$10$^{10}$ events/year in the energy range of interest (20 keVee to 60 keVee); 
this requires PSD to be 3.2 $\times$10$^{10}$ in order to achieve  a sensitivity of 3 $\times$ 10$^{-46}$ cm$^{2}$ 
at 20 keVee threshold. Such a high PSD has not been
yet demonstrated experimentally.  Moreover, without depletion of  $^{39}$Ar, it is unlikely that the data acquisition
 dead time would be manageable for a multi-ton 
detector and a large scale computing facility would be required to perform on-line data reduction for a large amount
 of data from $^{39}$Ar decays even for 
a single-phase detector. In addition, a dual-phase detector that drifts electrons from the liquid to 
gaseous would encounter serious difficulties 
with pile-up events because of the milliseconds drift time required.  With depletion of $^{39}$Ar by a large factor
 we can not only reduce background data, but also make 
it possible to lower the energy threshold
 to access a larger parameter space of WIMPs. Therefore, a multi-ton argon-based dark matter detector is promising only 
if natural argon is depleted of $^{39}$Ar. 

A large amount of argon depleted in $^{39}$Ar and $^{42}$Ar would be of great benefit to
 the DUSEL dark matter program described above.  
Because $^{39}$Ar and
$^{42}$Ar are produced by cosmic rays in the atmosphere, it is logical to consider that underground sources of argon
 should be depleted in $^{39}$Ar and $^{42}$Ar
due to the suppression of cosmic rays flux by the overburden.  There are two sources of underground argon: 1) argon  
in natural gas and 2) argon dissolved in underground water.
The former from natural gas wells in the US National Helium Reserve was found to contain a low level of $^{39}$Ar~\cite{dac}. 
The ratio of $^{39}$Ar to $^{40}$Ar (stable isotope) was found to be $\leq$ 4$\times$10$^{-17}$ at a confidence level of 84\%~\cite{dac}. 
The latter from underground water 3200 feet below the surface 
at Wall, South Dakota is the focus of this paper.  

Understanding the production of various argon isotopes underground by stopping negative muon captures and neutron interactions
 is critical to the dating method with $^{39}$Ar. In this study, we realize for the first time that 
the $^{39}$Ar production via stopping negative 
muon capture on $^{39}$K is
a dominant processes when the depth of underground resources is smaller than 1800 m.w.e. The early work of 
references~\cite{hhl, mcd} showed the $^{39}$Ar production only 
through (n,p) reaction on $^{39}$K.   
Neutrons can be produced by: 1)
cosmic-rays depending on the rock composition and depth of the site~\cite{mei}; and 2) ($\alpha, n)$ neutrons depending
 on the rock composition and radioactivity level in the
rock. The neutron fluxes and energy spectra are essential to the calculation of various isotopes of argon and hence 
the reliability of  dating results.
 In this paper, 
we describe the production mechanisms underground for argon content in Section~\ref{sec:prod}. The measured rock 
composition and 
potassium content in a site as a potential argon source
is discussed in Section~\ref{sec:rock}.  We show the production rate of $^{39}$Ar and $^{40}$Ar
in Section~\ref{sec:cal}. Finally,  we summarize our conclusions in Section~\ref{sec:con}. 
\section{The production mechanisms of argon isotopes underground}
\label{sec:prod}
\subsection{Production of $^{39}$Ar}
Underground production of $^{39}$Ar isotopes occurs primarily through  stopping negative muon captures and ($n,p$) reactions on
 $^{39}$K and $^{40}$Ca targets. Both elements are in a range of a few percent depending on the chemical composition 
of the rock.
Neutrons are produced by three sources: 1) ($\alpha$,n) reactions induced by $^{238}$U and $^{232}$Th decay chains;
 2) fission decays induced by $^{238}$U and $^{232}$Th; and 3) muon-induced neutrons. The production of $^{39}$Ar through 
stopping negative muon captures and muon-induced neutrons depends strongly on the depth of the rock overburden. 
This results in a depth dependent production of $^{39}$Ar 
induced by stopping negative muons and muon-induced neutrons. 
\subsubsection{Stopping negative muon capture on $^{39}$K}
When negatively charged muons are stopped in the rock, they either decay through
$\mu^{-}\rightarrow e^{-}\bar{\nu_e}\nu_{\mu}$
or are captured through $\mu(A, Z)\rightarrow \nu_{\mu}(A, Z-1)$.
The capture cross section can be estimated using the formula
\begin{equation}
\label{eq:cross}
\sigma = \frac{A}{N_{a}v\tau\rho},
\end{equation}
where A is atomic mass number of the isotope, 
$N_{a} = 6.022\times 10^{23}$ mol$^{-1}$ is Avogadro's constant, 
$v$ is the speed of the stopping negative muons in the medium, 
$\tau$ is the stopping negative muon capture time in the medium, 
and $\rho$ is the density of the medium.
The lifetime of stopping negative muons in potassium is 410 ns~\cite{tsu}. 
The capture time can be calculated based on
\begin{equation}
\label{eq:capt}
\frac{1}{\tau} = \frac{1}{\tau_{m}} - \frac{1}{\tau_{0}},
\end{equation}
where $\tau_{m}$ (410 {\it ns}) is the measured lifetime of muons in potassium and 
$\tau_{0}$ (2200 {\it ns}) is the lifetime of muons in vacuum. 
Using Eqs. (\ref{eq:cross}) and (\ref{eq:capt}), 
we obtain the cross section of the stopping negative muon capture on 
potassium as a function of the stopping muon energy.

The production rate of $^{39}$Ar per year per gram of  rock can be estimated using
\begin{equation}
\label{eq:pr}
P_{1} = \frac{\rho}{A} \cdot N_{a} \cdot \phi \cdot \sigma \cdot f_{neg} \cdot \eta,
\end{equation}
where P$_{1}$ is the production of $^{39}$Ar per year per gram of rock, $\rho$ is the density of potassium in the rock, 
$\phi$ is the stopping muon flux, $\sigma$ is the capture cross section, f$_{neg}$ = 0.44  is the fraction of 
negative muons~\cite{fuk}, and $\eta$ is the branching ratio of the yield of reaction $^{39}$K($\mu^{-}$,$\nu$)$^{39}$Ar. 
\subsubsection{Neutron-induced $^{39}$Ar production}
In the rock formed thousand years ago, the rates of production
and decay of $^{39}$Ar should be equal. The production rate 
per year per gram of rock can be calculated by
\begin{equation}
P_{2} = \phi_n \frac{N_{K}\sigma_{^{39}K(n,p)}}{N_{tot}\sigma_{abs}},
\label{eq:rate}
\end{equation}

\noindent
where $\phi_{n}$ is the neutron flux in neutrons per year per gram of rock, $N_K$ and $N_{tot}$ are the 
number of potassium atoms and the total number of atoms in a gram of rock, and  
$\sigma_{^{39}K(n,p)}$ and $\sigma_{abs}$ are the exothermic 
reaction $^{39}$K(n,p)$^{39}$Ar cross section and neutron absorption 
cross section. Note that the neutron absorption here means that the
neutron is absorbed and no further neutron is ejected, thus the
absorption cross section includes contributions from 
all reaction channels which do not eject neutrons. 

\begin{figure}[htb!!!]
\includegraphics[angle=0,width=8.cm] {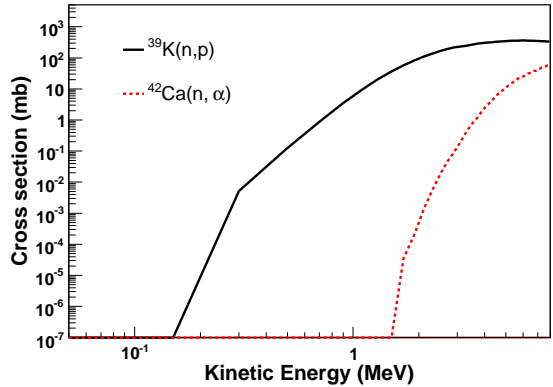}
\caption{\small{$^{39}$Ar production cross section as a function of neutron 
kinetic energy.}}
\label{fig:ar39}
\end{figure}

We calculate reaction cross sections by using the 
nuclear reaction code TALYS~\cite{talys} version 0.72.
The default input parameters are used in this work. 
Fig.~\ref{fig:ar39} shows the cross section of 
$^{39}$K(n,p) reaction as a function of neutron kinetic energy. 
Also shown in the figure
is the cross section of $^{42}$Ca(n,$\alpha)$ reaction, which can
in principle contribute to $^{39}$Ar production. However, its contribution
is negligible because of relatively small cross section and 
$^{42}$Ca concentration compared to $^{39}$K.

\begin{figure}[htb!!!]
\includegraphics[angle=0,width=8.cm] {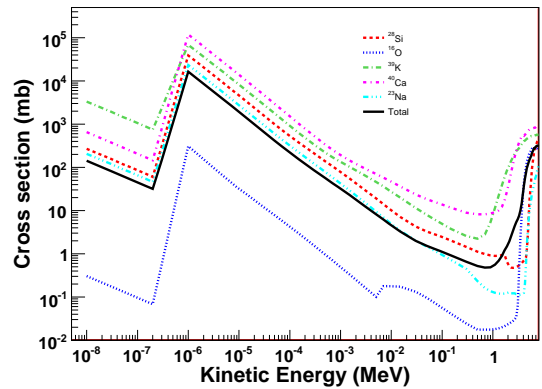}
\caption{\small{The cross section of neutron absorption in different
nuclei as a function of neutron kinetic energy. The black line 
shows neutron absorption cross section in the rock.}}
\label{fig:absXs}
\end{figure}

We plot in Fig.~\ref{fig:absXs} the neutron absorption cross section of 
major atomic components in the rock. The black line illustrates the 
total neutron absorption cross section that is the weighted 
average, which uses the number of atoms as the weight factor. 

The number of $^{39}$Ar atoms is given by 
\begin{equation}
\label{eq:half}
N_{^{39}Ar} = \frac{P_{1} + P_{2}}{\lambda},
\end{equation}

\noindent
where P$_{1}$ and P$_{2}$ are the $^{39}$Ar production rates given by Eq.(\ref{eq:pr}) and Eq.(\ref{eq:rate}), 
$\lambda = \ln 2/t_{1/2}$ is the decay constant of $^{39}$Ar.
The half-life of $^{39}$Ar is 269 years. 

\subsection{The production rate of $^{40}$Ar}
\begin{figure}[htb!!!]
\includegraphics[angle=0,width=8.cm] {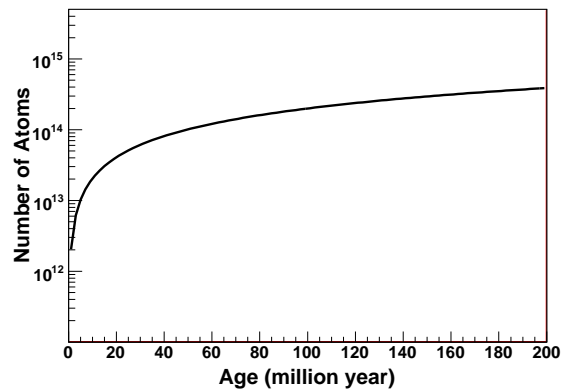}
\caption{\small{The number of $^{40}$Ar atoms produced in a gram of rock
as function of the rock age.}}
\label{fig:ar40}
\end{figure}

$^{40}$Ar in rock is produced  mainly by the decay of $^{40}$K 
through electron capture with a branch ratio of 10.72\%. 
The number of $^{40}$Ar atoms produced can be estimated according to
\begin{equation}
\label{eq:ar}
N = N_0 B (1-\exp(-\lambda t)),
\end{equation}

\noindent
where $N_0$ is original $^{40}$K atoms at time $t = 0$, $B$ is the branch 
ratio of electron capture, and $\lambda = \ln 2/t_{1/2}$ is the decay constant.
The half-life $t_{1/2}$ of $^{40}$K is $1.277\times 10^{9}$ years. 
Fig.~\ref{fig:ar40} shows the number of $^{40}$Ar atoms 
produced as a function of the rock age. Although the natural 
abundance of $^{40}$K is only 0.0117\%, the number
of $^{40}$Ar produced in 65 million years reaches $\sim1.7\times 10^{14}$ 
atoms per gram of rock assuming a 2.59\% of potassium content in the rock. 

Stopping negative muon capture on $^{40}$Ca produces $^{40}$K, which generates $^{40}$Ar through electron capture decay. Because the
production rate is much smaller than the existing $^{40}$K in the rock, we ignore this process in our calculation of 
the ratio of $^{39}$Ar to $^{40}$Ar.

\subsection{Calculation of the $^{39}$Ar/$^{40}$Ar ratio as a function of depth}

We utilize the fluxes of stopping negative muons and muon-induced neutrons as a function of depth derived in Ref.~\cite{mei} 
to calculate 
the production of argon content with equations (\ref{eq:pr},\ref{eq:rate},\ref{eq:half},\ref{eq:ar}). The neutron energy
spectrum induced by radioactivity obtained in Ref.~\cite{meichao} is used in the calculation. 

The common elements found in the Earth's rocks are used in the calculation~\cite{pid}. 
The eight most common elements~\cite{pid} used in the calculation are listed in Table~\ref{tab:rock}. The natural
 radioactivity concentrations of $^{238}$U and $^{232}$Th depend on the rock types. We assume the common rock 
in Table~\ref{tab:rock} is granite. Correspondingly, the radioactivity level of 4.7 ppm and 2.0 ppm 
of $^{238}$U and $^{232}$Th are used in the calculation~\cite{rock}. Using the method described in Ref.~\cite{meichao}, 
we obtain
a yield of 5.5 n/(g$\cdot$y). Subsequently, the $^{39}$Ar production rate is estimated to be 7$\times$10$^{-3}$ atoms per year per 
gram of rock. The total number of $^{39}$Ar atoms is 2.7 per gram of rock, which is independent of depth.

\begin{table}[htb!!!]
\caption{A list of the eight most common elements found in the Earth's rocks.}
\begin{tabular}{lll}
\hline \hline
Elements&Chemical Symbol & Mass Weight (\%) \\
\hline
Oxygen&O & 46.60\\
Silicon&Si&27.72\\
Aluminum&Al&8.14\\
Iron&Fe&5.00\\
Calcium&Ca&3.63\\
Sodium&Na&2.83\\
Potassium&K&2.59\\
Magnesium&Mg&2.09\\
\hline \hline
\end{tabular}
\label{tab:rock}
\end{table}

\begin{figure}[htb!!!]
\includegraphics[angle=0,width=8.cm] {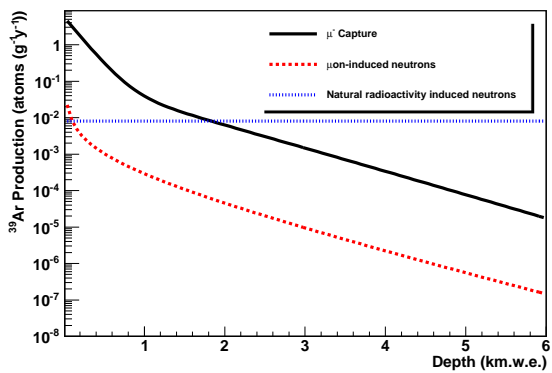}
\caption{\small{The $^{39}$Ar production as a function of depth. The eight common elements found in the Earth's rocks are 
 used in the calculation.}}
\label{fig:muons}
\end{figure}

The depth dependence of the $^{39}$Ar production is shown in Figure~\ref{fig:muons}. Note that the rock chemical 
 composition varies 
from location to location. However, the dependence of the production of $^{39}$Ar due to the change in the rock
chemical composition is much weaker than that of the change in depth.

\begin{figure}[htb!!!]
\includegraphics[angle=0,width=8.cm] {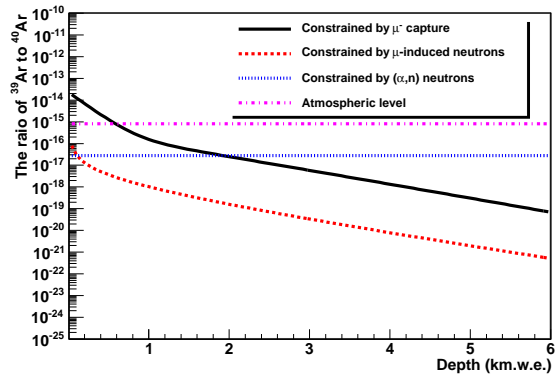}
\caption{\small{The ratio of the $^{39}$Ar to $^{40}$Ar calculated using formulas and methods described in text. 
The eight common elements found in the Earth's rocks are 
 used in the calculation. Shown is a ratio assuming that the produced argon 
is mixed at a 10\% level by underground gas. }}
\label{fig:muons1}
\end{figure}

If one assumes that $^{39}$Ar and $^{40}$Ar have the same
possibility to diffuse out of the grains of the rock and to be mixed with existing gas in
the pore volume, then the $^{39}$Ar/$^{40}$Ar ratio in the pore volume can be expressed as
\begin{equation}
\label{eq:ra}
R = \frac{ N_{^{39}Ar} \times p}{N_{Ar}^{0} + N_{Ar}^{d} \times p}
\end{equation}

\noindent
where $p$ is the possibility of $^{39}$Ar and $^{40}$Ar being mixed in the pore volume, N$_{Ar}^{0}$ is 
the original number of $^{40}$Ar contained in 
a certain amount of gas due to the $^{40}$Ar concentration in ambient air, which is related to the pore volume, N$_{Ar}^{d}$ is the number of $^{40}$Ar atoms
produced underground through $^{40}$K EC decays. Using Eq.~(\ref{eq:ra}), the ratio of the $^{39}$Ar to $^{40}$Ar 
can be calculated with known natural abundance of $^{39}$K in the rock.
Given a rock density of 2.7 g/cm$^{3}$, a rock porosity of 20\%, and an 
assumption that the pore
volume is filled with gas with the ambient
air concentrations (78\%/21\%/1\% of N$_{2}$/O$_{2}$/Ar), the ratio of $^{39}$Ar to $^{40}$Ar can be determined.  
The results are shown in Figure~\ref{fig:muons1}. 

\begin{figure}[htb!!!]
\includegraphics[angle=0,width=8.cm] {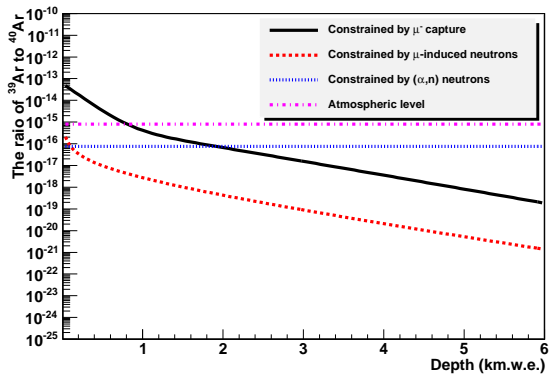}
\caption{\small{The ratio of the $^{39}$Ar to $^{40}$Ar calculated using formulas and methods described in text. 
The eight common elements found in the Earth's rocks are 
 used in the calculation. Shown is a ratio assuming that the produced argon 
is dissolved at a 10\% level by underground water. }}
\label{fig:muons2}
\end{figure}

Eq.(\ref{eq:ra}) can be applied to a underground water reservoir, where $p$ is the possibility of $^{39}$Ar and
 $^{40}$Ar being dissolved in water.  For a rock density of 2.7 g/cm$^{3}$, a rock porosity of 20\%, and an 
assumption that the pore water contains gas with the ambient
air concentrations (78\%/21\%/1\% of N$_{2}$/O$_{2}$/Ar) at a solubility of 61 mg/L for argon, the ratio
 of $^{39}$Ar to $^{40}$Ar as a function of depth is shown in Figure~\ref{fig:muons2}. 

As can be seen in Fig.~\ref{fig:muons}, the larger the depth, the smaller the production of $^{39}$Ar. On the other hand, 
at the large depth, the production of $^{39}$Ar is dominated by stopping negative muon capture and the neutrons 
from natural radioactivity. Since $^{40}$Ca$(n,2p)$$^{39}$Ar reaction requires neutron energy to be greater than 8.3 MeV and 
the majority of the $(\alpha,n)$ neutrons induced by radioactivity is less than this reaction threshold, 
the production of $^{39}$Ar at large depth underground is dominated by stopping negative muon capture on $^{39}$K and 
$^{39}$K(n,p)$^{39}$Ar. Hence, the understanding of $^{39}$K content and radioactivity level in the rock surrounding 
the gas field or water reservoir where the argon gas is extracted from is essential.  

The underground production of $^{40}$Ar is primarily from $^{40}$K via electron capture decays. Therefore, an accurate 
knowledge  of the potassium content is critical. The underground production of $^{42}$Ar is negligible. 

\section{Experimental results of potassium content}
\label{sec:rock}
\subsection{Underground water Site}
The community of Wall, South Dakota pumps water using seven wells that extract water from the Fall River 
aquifer in an old rock formation 3200 ft below the surface. 
The average water consumption is 200 gallons per minute. The water temperature underground is about 42$^{\circ}$C.
The expected argon gas extraction is about 2.3 kg per day assuming 70\% efficiency with 61 mg/L solubility. However,
the level of  $^{39}$Ar in 
argon gas extracted from well water is contingent upon the understanding of the rock composition especially the potassium
 content in the rock. Two different samples obtained from the South Dakota 
Geological Survey were evaluated. The first sample was a cutting sample from the 3200 ft level in the well that was 
provided by the well drilling company. The second sample was a core sample taken from the Fall River Aquifer at a distance of 
about 40 miles from Wall.
\subsection{Method}
The  samples  were  broken  down  using  a  mortar and pestle  and  then  passed  through a 100 mesh  sieve.  The samples were then dried in an oven set at 100$^{\circ}$C to remove the water.  After drying the samples were consistently stored in a desiccator to prevent re-hydration. Initially  the  amount  of free  K  (K  not  in  the  soil matrix)  was found using the  following solution  preparation and  utilizing  the  atomic  absorption spectrometer.  The solution  preparation consisted  of 2 g of the  sample,  40 ml of water  (Note that all water used was nanopure  water  that had  been  purified  to  18 M$\Omega$ cm with a SYBRON/Barnsted Nanopure System), 15.5 ml of concentrated hydrochloric  acid (HCl),  and  2 ml of concentrated nitric  acid  (HNO$_{3}$ ).   One  important observation  was that the  solution  did  not  boil when  the  acids were added  indicating  a lack of carbonates.  The solution was heated with stirring, and finally filtered to remove the remaining soil particles.    The  filtered  solution  was placed  in a 250 ml volumetric  flask and  diluted  with  water  to  the  250 ml mark.   An atomic absorption analysis was done on this sample using a Thermo Jarrell Ash atomic absorption spectrometer which contained a Smith-Hieftje 11 spectrophotometer.  The light source was a hollow cathode lamp produced by the Thermo Jarrell Ash Corporation which emitted at 766.5 nm.  The bandpass was set at 1.0 nm and the lamp current was set to 5 mA.  Two runs were done with the instrument for each sample.
The measurement of the total K in the soil samples was measured under the assumption that most of the K was tied up in a SiO$_{2}$   compound.   This was concluded from the X-ray diffraction analysis conducted on the sample. The sample preparation was obtained from Buck Scientific~\cite{htt}.   One gram of the soil sample was mixed with 5 grams of a 2:1 flux of lithium carbonate (LiCO$_{3}$) and zinc oxide (ZnO).  The sample and flux were placed in a platinum crucible and fused in a muffle furnace that was heated to 950$^{\circ}$C for 20 minutes.   The melt was then allowed to cool and was re-dissolved using 20 ml of HCl and 20 ml of water.  This solution was filtered to remove the soil particles that had not been re-dissolved.  A series of standards that ranged from 0.01 ppm to 5.0 ppm were prepared and analyzed with the atomic absorption spectrophotometer yielding a calibration curve of $y = (0.0562 \pm 0.0009) x - (0.0034 \pm$ 0.0021) with a correlation coefficient of 0.9994. Potassium samples were diluted to fall within the calibration curve. Once again two runs were done and then averaged.   Three different samples from the core sample and from the cutting sample were treated in this manner. Finally, we obtain
a potassium content of 18500 $\pm$ 300 ppm. 

\section{Production rate of argon isotopes}
\label{sec:cal}

 In this particular site in
South Dakota at 3200-ft below the surface,  the stopping muon flux is calculated to be
3.4$\times$10$^{-8}$cm$^{-2}$s$^{-1}$. Utilizing Eq.(\ref{eq:pr}), we obtain a $^{39}$Ar production of 0.0026 per year per 
gram of rock.

The natural radioactivity of $^{232}$Th 
and $^{238}$U is analyzed to be around 
0.6 ppm and 1.9 ppm, separately. The rock is sandstone with a chemical composition described in Table~\ref{tab:sandstone}.
\begin{table}[htb!!!]
\caption{The chemical composition of the sandstone 3200 ft below the surface at Wall, South Dakota.}
\begin{tabular}{ll}
\hline \hline
Elements & Mass Weight (\%) \\
\hline
Si& 43.0\\
O&50.71\\
K&1.85\\
Fe&1.5\\
Kr&1.0\\
Na&0.7\\
Mg&0.6\\
Al&0.5\\
Ca&0.14\\
\hline \hline
\end{tabular}
\label{tab:sandstone}
\end{table}

The ($\alpha,n)$ neutron yield in the rock is calculated~\cite{meichao} to be 0.25 n/(g$\cdot$ppm$\cdot$y) for $^{232}$Th and 
0.65 n/(g$\cdot$ppm$\cdot$y) for $^{238}$U.
The fission decay of $^{238}$U induced neutron yield is 0.52 n/(g$\cdot$ppm$\cdot$y). The total neutron yield induced by radioactivity is
2.4 n/(g$\cdot$y). Using Eq.(\ref{eq:rate}), the $^{39}$Ar production rate is estimated to be $\sim$3$\times$10$^{-3}$ atoms per year 
per gram of  rock. 
 
The $^{39}$Ar produced by the muon-induced neutrons at this depth is negligible. The total number of $^{39}$Ar atoms
given by Eq.(\ref{eq:half}) is estimated to be 2.17 atoms per gram of rock.

\begin{figure}[htb!!!]
\includegraphics[angle=0,width=8.cm] {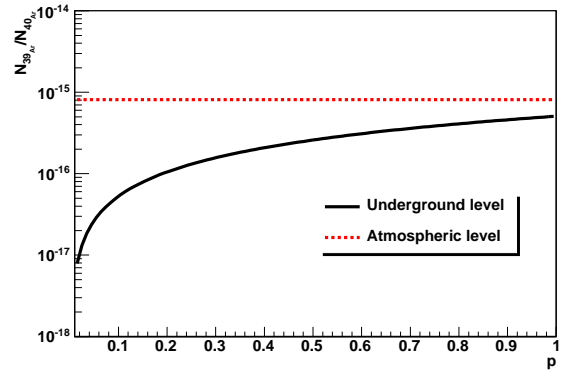}
\caption{\small{The ratio of $^{39}$Ar to $^{40}$Ar atoms 
contained in water as a function of the dissolved probability ($p$).}}
\label{fig:ratio}
\end{figure}

If we assume that $^{39}$Ar and $^{40}$Ar have the same
possibility to diffuse out of the grains of the rock and to be dissolved in
the pore water, then the $^{39}$Ar/$^{40}$Ar ratio in water can be obtained by applying Eq.~(\ref{eq:ra}),
\begin{equation}
R = \frac{2.17 p}{N_{Ar} + 1.2\times 10^{14} p}
\end{equation}

\noindent
where $p$ is the possibility of $^{39}$Ar and $^{40}$Ar being dissolved in
water, and $N_{Ar}$ is the original number of $^{40}$Ar contained in 
a certain amount of water, which is related to the pore volume. If we
assume that the porosity is 20\% (this is typical for sandstone~\cite{hhl}), then the water volume corresponding 
to a gram of rock with the measured density of 2.2 g/cm$^3$ is $9.1\times 10^{-2}$ cm$^{3}$. Given 
 a total saturation solubility of 125 mg/L  for N$_{2}$/O$_{2}$/Ar at 20$^{\circ}$C assuming the gas has ambient air 
concentrations (nitrogen/oxygen/argon of 78\%/21\%/1\%) with different solubility of N$_{2}$ (20 mg/L)/O$_{2}$ 
(44 mg/L)/ Ar (61 mg/L), the dissolved argon in water is 
61 mg/L $\times$ 0.01 / (20 mg/L $\times$ 0.78 + 44 mg/L $\times$ 0.21 + 61 mg/L $\times$ 0.01).  
This results in a 2.4 \% of argon content ( 3 mg/L)
in the total gas content in the water. Therefore, $N_{Ar} = 4.1\times 10^{15}$.
The term of 1.2$\times$10$^{14}$ is the number of 
$^{40}$Ar atoms from $^{40}$K decay with a total concentration of 18500 ppm in a $\sim$65 million-year rock formation. 
Fig.~\ref{fig:ratio}
shows the ratio of $^{39}$Ar to $^{40}$Ar atoms contained in
water as a function of $p$. 
Thus, a depletion factor of 1.6 to 155 can be achieved for the
argon extracted from this site when p varies from 0.01 to 1.0.
Considering saturation solubility decreases when temperature increases, the dissolved possibility of  
$^{39}$Ar and $^{40}$Ar can be as small as 1\% for the argon content produced in the underground environment where 
the temperature is higher. This would make underground water reservoir a valuable source for depleted argon that can be used for ton-scale dark matter experiments. 
 If one assumes that 30\%~\cite{hhl} of Ar produced 
will be dissolved in water, the ratio between $^{39}$Ar and $^{40}$Ar
in water is $\sim 1.6\times 10^{-17}$, which is a factor of $\sim$5
lower than that in atmosphere. 

\section{Conclusion}
\label{sec:con}
 We have evaluated that the $^{39}$Ar production underground through $(n,p)$ reactions induced by neutrons and stopping negative muon capture. We demonstrate that argon depleted in $^{39}$ can be obtained only if the depth of the underground resources is greater than 500 m.w.e. below the surface. Stopping negative muon capture on $^{39}$K dominates over radiogenic production at depths of less than 2000 m.w.e., and that production by muon-induced neutrons is subdominant at any depth. Depending on the rock porosity and argon
diffusion coefficient, the upper limit of the concentration of $^{39}$Ar in the underground water at Wall of South Dakota is estimated to be in a range of a factor of 1.6 to 155 less 
than the $^{39}$Ar concentration in the atmosphere. 
Naturally occurring $^{40}$K EC decay dominates the content of $^{40}$Ar underground.
The ratio of $^{39}$Ar to $^{40}$Ar depends largely on the following: 1) the potassium content in the rock; 2) the natural radioactivity of $^{232}$Th and $^{238}$U 
in the rock; 3) the rock porosity; 4) argon gas diffusion coefficient underground;  and 5) argon gas solubility in the water underground.
We estimate that the production of $^{42}$Ar is negligible at this underground site. The calculation tools presented in this paper are also critical to the
dating method with $^{39}$Ar~\cite{hhl, hhh, mcd, pwz}. We conclude that the ratio of $^{39}$Ar to $^{40}$Ar varies from location to location depending on the $^{39}$K content and the radioactivity level in the rock at large depth. Extraction of  argon gas from underground sources must: 1)  evaluate the depth of the site, the age of the rock, the rock composition, and radioactivity; 2) study the porosity of the rock and argon gas diffusion coefficient; and 3) understand the solubility of argon gas as a function of pressure and temperature.  The algorithm described in this paper is critical to understanding the discrepancy in the dating methods with the ratio of $^{39}$Ar to $^{40}$Ar~\cite{mcd, iwe, kom, fjp}. 
\section*{Acknowledgments}
The authors wish to thank to Cristiano Galbiati for a careful reading of this manuscript. 
This work was supported in part by NSF grant PHY-0758120 and the Office of Research at University 
of South Dakota as well as  by Laboratory Directed Research 
and Development at Los Alamos National Laboratory. J. Spaans was also partially supported by the South Dakota Space Grant Consortium.  Z.Y. 
was also partly supported by MOE of China under project No. IRT0624 
and the NSFC under grant No. 10635020 and 10975061.

%
%

\begin{thebibliography}{99}
\bibitem{bh} M. G. Boulay and A. Hime, Astropart. Phys. {\bf 25}, 179 (2006).
\bibitem{warp}A. Rubbia, J. Phys. Conf. Ser. {\bf 39}, 129 (2006); 
Benetti {\it et al.}, Nucl. Instr. Meth. A {\bf 574}, 1 (2007).
\bibitem{hhl} H. H. Loosli, Earth Plan. Sci. Lett. {\bf 63}, 51 (1983).
\bibitem{wku} W. Kutschera {\it et al.}, Nucl. Instr. Meth. B {\bf 92}, 241 (1994).
\bibitem{wrap1} Benetti {\it et al.} (WARP Collaboration), Nucl. Instr. Meth. A {\bf 574}, 83 (2007).
\bibitem{sto} R. W. Stoenner, O. A. Schaeffer, S. Katcoff, Science 148/3675: 1325 (1965).
\bibitem{car} C. Arpesella, C, Brofferio, P. P. Sverzellati, M. Zhou and F. Cavanna, Preprint LNGS 92/27 (1992).
\bibitem{pce} P. Cennini {\it et al.}, Nucl. Instr. Meth. A {\bf 356}, 526 (1995).
\bibitem{asb} A. S. Barabash, V. N. Kornoukhpv, and V. E. Jants, Nucl. Instr. Meth. A {\bf 385}, 530 (1997).
\bibitem{ahi} A. Hitachi {\it et al.}, Phys. Rev. B {\bf 27}, 5279 (1983). 
\bibitem{pbe} P. Benetti {\it et al.} (WARP Collaboration), in press on Astropar. Phys., arXiv:astro-ph/0701286.
\bibitem{hugh} W. H. Lippincott {\it et al.}, arXiv:nucl-ex/0801.1531; Phys. Rev. C {\bf 78}, 035801 (2008).
\bibitem{dac} D. Acosta-Kane {\it et al.} (WARP Collaboration), arXiv: astro-ph/0712.0381, Nucl. Instr. Meth. A {\bf 587} 46 (2008).  
\bibitem{mcd} Ian McDougall and T. Mark Harrison, ``Geochronology and Thermochronology by the $^{40}$Ar/$^{39}$Ar Method'', Oxford University Press, USA (1999).
\bibitem{mei} D.-M. Mei and A. Hime, Phys. Rev. D {\bf 73}, 053004 (2006).
\bibitem{tsu} T. Suzuki, D. F. Measday, J. P. Roalsvig, Phys. Rev. {\bf C 35}, 2212 (1987).
\bibitem{fuk} Fukahori, T., Proc. of the Specialists' Meeting on High Energy Nuclear Data, Tokai, October 3-4, 1991, JAERI-M 92-039, pp. 114-122 (1992).
\bibitem{talys} A. J. Koning, S. Hilaire and M. C. Duijvestijn, 
``TALYS: Comprehensive nuclear reaction modeling'', Proceedings of the
International Conference on Nuclear Data for Science and Technology - 
ND2004, AIP vol. 769, eds. R. C. Haight, M. B. Chadwick, T. Kawano,
and P. Talou, Sep. 26-Oct. 1, 2004, Sante Fe, USA, p. 1154 (2005).
\bibitem{meichao} D.-M. Mei, C. Zhang, A. Hime, Nucl. Instr. Meth. A {\bf 606}, 651 (2009). http://neutronyield.usd.edu.
\bibitem{pid} M. Pidwirny, ``Composition of Rocks.'' Fundamentals of Physical Geography, 2nd Edition (2006), Date Viewed. 
http://www.physicalgeography.net/fundamentals/10d.html.
\bibitem{rock} M. Eisenbud and T. Gesell, Environmental Radioactivity from Natural, Industrial and Military Sources, 4$^{th}$
Edition, Academic Press, Inc.
\bibitem{htt} G. J. DeMenna, http://www.bucksci.com/PDF/aa3001.pdf.
\bibitem{hhh} H. H. Loosli and H. Oeschger, Isotope Hydrology 1978, 2 (IAEA, Vienna, 1979) 931-947; H. H. Loosli and H. Oeschger, Radiocarbon {\bf 22}, 863 (1980).
\bibitem{pwz} P. Collon, W. Kutschera, Z.-T. Lu, Ann. Rev. Nucl. Part. Sci., Vol. {\bf 54}, 39 (2004); Nucl-ex/040213.
\bibitem{iwe} I. Wendt, M. A. Geyh and F. Fauth, Isotope Hydrology 1967 (IAEA, Vienna, 1967) 321-337.
\bibitem{kom} K. O. Munnich, Isotope Hydrology 1970 (IAEA, Vienna, 1970) 259-270.
\bibitem{fjp} F.J. Pearson and B. B. Hanshaw, Isotope Hydrology 1970 (IAEA, Vienna, 1970) 271-286.
\end{thebibliography}

\end{document}